\documentstyle[aps,multicol,prl]{revtex}

\begin{document}
\bibliographystyle{prsty}

\title{Lattice dynamics of BaTiO$_3$, PbTiO$_3$ and PbZrO$_3$:\\
a comparative first-principles study}
\author{Ph. Ghosez, E. Cockayne\cite{*}, 
U. V. Waghmare\cite{**} 
and K. M. Rabe}
\address{Department of Applied Physics, Yale University, 
P.O. Box 208284, New Haven, CT 06520-8284, USA.}
\date{\today}
\maketitle
\begin{abstract}
The full phonon dispersion relations of lead titanate 
and lead zirconate in the cubic perovskite structure
are computed using first-principles variational density-functional perturbation  
theory, with ab initio pseudopotentials and a plane-wave basis set.
Comparison with the results previously obtained for barium titanate shows 
that the change of a single constituent  
(Ba to Pb, Ti to Zr) has profound effects on the character and dispersion of  
unstable modes,  with significant implications for the nature of the phase  
transitions and the dielectric and piezoelectric responses of the compounds.
Examination of the interatomic force constants in real space, obtained by a  
transformation which correctly treats the long-range dipolar contribution,  
shows that most are strikingly similar, while it is the differences in a few  
key interactions which produce  the observed changes in the  
phonon dispersions.  
These trends suggest the possibility of the transferability of force constants  
to predict the lattice dynamics of perovskite solid solutions.
\end{abstract}

\begin{multicols}{2}[]

\setcounter{page}{1}

\section{Introduction}

In the family of  
perovskite ABO$_3$ compounds, a wide variety of distorted variants of the 
high-temperature cubic structure are observed as a function of composition 
and temperature\cite{Lines77}. 
First-principles density-functional energy-minimization methods have proved 
to be generally quite accurate in the theoretical prediction of the ground 
state structure type and structural parameters of perovskite oxides\cite{Vanderbilt97}.
According to the soft-mode theory of structural phase transitions, the 
ferroelectric phases can be related to the high temperature symmetric structure 
by the freezing-in of unstable zone-center phonons.
However, to predict finite temperature behavior at phase transitions, as well 
as the temperature dependence of the dielectric and piezoelectric responses 
of the compounds, it is necessary to have information about the energies of 
{\it non-uniform} instabilities and low energy distortions of the cubic 
perovskite structure~\cite{RW92}.
For this purpose, the calculation of the phonon dispersion relations through 
density-functional perturbation theory (DFPT)~\cite{DFPT,Gonze92} has been 
found to be extremely useful, allowing easy identification of the unstable phonon 
branches and their dispersion throughout the Brillouin zone.
This information also permits the investigation of the geometry of localized 
instabilities in real space, which can be directly related to the anisotropy 
of the instability region in reciprocal space~\cite{Yu95b,Rabe98,Ghosez98a}.
To date, phonon calculations away from the zone center have been reported for 
only a few individual compounds, with full phonon dispersion relations given for 
KNbO$_3$~\cite{Yu95b}, SrTiO$_3$~\cite{LaSota97} and BaTiO$_3$~\cite{Ghosez98a}, 
and selected eigenmodes for PbTiO$_3$~\cite{Waghmare97a}.

In this paper, we present the first comparative study of the full phonon dispersion relations 
of three different cubic perovskites: BaTiO$_3$, PbTiO$_3$ and PbZrO$_3$,
computed from first principles using the variational DFPT method as described in Section II.
These compounds have been chosen both because of their scientific and technological importance, 
and because they allow us to investigate the effects on the lattice dynamics of substituting 
one cation with the other atoms unchanged.
As we will see in Sections III and IV, these substitutions lead to very pronounced differences 
in the eigenvectors and dispersions of the unstable phonons. 
The origin of these differences will be clarified in Section V, where, using a systematic 
method previously applied to BaTiO$_3$~\cite{Ghosez98a}, the reciprocal space force constant 
matrices are transformed to obtain real-space interatomic force constants for all three compounds.
There, we will see that the differences can be attributed to trends in cation-oxygen 
interactions, with other IFCs remarkably similar among the three compounds. 
These trends and similarities, and their implications for the study of the lattice dynamics 
of solid solutions, are discussed in Section V.
Section VI concludes the paper.

\section{Method}

The first-principles calculations for BaTiO$_3$ follow the method 
previously reported in Ref.~\onlinecite{Ghosez98a}.
For PbTiO$_3$ and PbZrO$_3$, calculations were performed within the Kohn-Sham formulation 
of density functional theory~\cite{DFT}, using the conjugate-gradients method~\cite{RMP}. 
The exchange-correlation energy 
functional was evaluated within the local density approximation (LDA), using 
the Perdew-Zunger parametrization~\cite{Perdew81} of the 
Ceperley-Alder~\cite{Ceperley80} homogeneous electron gas data~\cite{Rem-XC}. 
The ``all-electron'' potentials were replaced by the same {\it ab initio} 
pseudo-potentials 
as in Refs. \onlinecite{Waghmare97a} and \onlinecite{Waghmare97b}. 
The electronic wavefunctions were expanded 
in plane waves up to a kinetic energy cutoff of 850 eV. Integrals over the  
Brillouin zone were approximated by sums on a $4 \times 4 \times 4$ mesh of 
special $k$-points~\cite{Monkhorst76+}.

The optical dielectric constant, the Born effective charges and the force  
constant matrix at selected $q$-points of the Brillouin zone were computed 
within a variational formulation~\cite{Gonze92} of density functional 
perturbation theory~\cite{DFPT}. The phonon dispersion curves were 
interpolated following the scheme  described 
in Ref.~\cite{Giannozzi91,Gonze94}. In this approach, the long-range character 
of the dipole-dipole contribution is correctly handled by
first subtracting it from the force constant matrix in reciprocal space and  
treating it separately. The short-range contribution to the interatomic 
force constants in real space is then obtained from the 
remainder of the force constant matrix in $q$-space using a discrete Fourier 
transformation. In this work, 
the short-range contribution was computed from the force constant matrices 
on a $2 \times 2 \times 2$ centered cubic mesh of $q$-points comprised of  
$\Gamma$, X, M, R and the $\Lambda$ point halfway from $\Gamma$ to 
R~\cite{Note-mesh}.  
{}From the resulting set of interatomic force constants in real space, the 
phonon spectrum can be readily obtained at any point in the Brillouin zone.

Our calculations have been performed in the cubic perovskite structure. For 
PbTiO$_3$, the optimized LDA lattice parameter (3.883 \AA) slightly 
underestimates the experimental estimated value of 3.969 \AA~\cite{PTa0} 
and we have decided 
to work, as for BaTiO$_3$~\cite{Ghosez98a}, at the experimental volume. For 
PbZrO$_3$, we have chosen, as in Ref. 
\onlinecite{Singh95}, to work at the optimized lattice parameter 
of 4.12 \AA, which is nearly indistinguishable from the extrapolated experimental 
value of 4.13 \AA~\cite{PZa0}.

\section{Dielectric properties}

Knowledge of the Born effective charges ($Z_\kappa^*$) 
and the optical dielectric tensor ($\epsilon_{\infty}$) is essential  
for describing the long-range dipolar contribution to the lattice dynamics 
of a polar insulator. In Table~\ref{Table.Z}, we present results for PbTiO$_3$ 
and PbZrO$_3$, computed using the method described in Section II, and for 
BaTiO$_3$, obtained previously in Ref.~\cite{Ghosez98a}.
The effective charges have been corrected 
following the scheme proposed in Ref.~\cite{Gonze97b} in order to satisfy
the charge neutrality sum rule. Our results differ by at most 0.09 electrons
with values reported for cubic PbTiO$_3$ and PbZrO$_3$ using slightly different 
methods and/or lattice constants~\cite{Waghmare97a,Waghmare97b,Zhong94}.

As usual  in the class of perovskite ABO$_3$ compounds, the amplitudes of some elements 
of the effective charge tensors deviate substantially from the nominal value 
expected in a purely ionic picture (for a review, see for 
instance Ref.~\cite{Ghosez98b}). 
This effect is especially pronounced for the Ti and associated O$_{\parallel}$ charges 
in BaTiO$_3$ and PbTiO$_3$. It reflects the sensitivity to atomic
displacement of the partially-covalent character of the Ti--O bond~\cite{Michele}.
In contrast, the effective charge of the Zr atom in PbZrO$_3$ and the  
associated O$_{\parallel}$ 
charge are significantly closer to their nominal ionic values
of $+4$ and $-2$, respectively.  The Zr effective
charge is comparable to that reported recently
for ZrO$_2$ ($+5.75$)~\cite{Detraux98}. 
In addition, 
in both PbTiO$_3$ and PbZrO$_3$, the {\it anomalous} contribution to the Pb  
charge, beyond the nominal charge of +2, is more than twice as large as that 
for the Ba charge in BaTiO$_3$. This feature, too, with a concomitent increase  
in the magnitude of $Z^*_{O{\perp}}$, reflects the sensitivity to atomic 
displacement of the partially covalent character of the bond between lead and 
oxygen.

Within the LDA, the computed optical dielectric constant (Table~\ref{Table.Z}) 
usually overestimates the experimental value.  The error is of the order of
20\% in BaTiO$_3$~\cite{Ghosez95a}, for which the extrapolated experimental 
value is 5.40~\cite{BTepsilon}, consistent with analogous comparisons in
KNbO$_3$~\cite{Wang96b} and SrTiO$_3$~\cite{LaSota97}. For PbZrO$_3$, 
it appears that experimental data is available only for the orthorhombic
phase, where the value is about 4.8~\cite{Jona55}, significantly less than the 
value of 6.97 which we have computed for the cubic phase. For PbTiO$_3$, 
our value of 8.24 is comparable to a recent first-principles result of 8.28 
obtained using a different method~\cite{BernardiniU}. In contrast to the other
perovskites, this represents a slight underestimate of the extrapolated 
experimental value of 8.64 reported in Ref. \cite{Zhong94}.

The origin of the LDA error in the optical dielectric constant
is a complex question. It arises at least partly from the lack of polarization
dependence of the approximate exchange-correlation functional~\cite{Gonze95}.
In the cubic phase of the perovskite ferroelectrics, the comparison with 
experiment is also complicated by the fact that the high-temperature cubic 
phases for which the measurements are made do not have a perfect cubic perovskite 
structure, as assumed in the calculations. In fact, the observed cubic structure 
of BaTiO$_3$ and PbTiO$_3$ represents the average of large local distortions.
The character of these distortions depends strongly on the material, which could 
well have different effects on the observed optical dielectric constant.

For the BaTiO$_3$ phonon dispersion, it has been checked that the inaccuracy 
in $\epsilon_{\infty}$ only significantly affects the position of the highest  
longitudinal optic branch, while other frequencies are relatively insensitive 
to the amplitude of the dielectric constant~\cite{Ghosez98a}. The effects of 
possible discrepancies for the other two compounds are likely to be 
similarly minor.

\section{Phonon dispersion curves}

In this section, we describe phonon dispersion relations for  
BaTiO$_3$, PbTiO$_3$ and PbZrO$_3$, providing a global 
view of the quadratic-order energy surface around the cubic perovskite 
structure. The calculated phonon dispersion curves along the high symmetry 
lines of the simple cubic Brillouin zone are shown in Fig.~\ref{Fig.BS}. 
The unstable modes, which determine the nature of the phase transitions and 
the dielectric and piezoelectric responses of the compounds, have imaginary  
frequencies. Their dispersion is shown below the zero-frequency line. 
The character of these modes also has significant implications for the
properties of the system. 
This character has been depicted in Fig.~\ref{Fig.BS} by assigning a color to each  
eigenvalue, determined by the percentage of each atomic character in the 
normalized eigenvector of the dynamical matrix (red for A atom, green for 
B atom and blue for O atoms)~\cite{Note-color}.

Barium titanate (BaTiO$_3$) and potassium  
niobate (KNbO$_3$) both undergo a transition sequence with decreasing temperature 
through ferroelectric tetragonal, orthorhombic and rhombohedral (ground state) 
structures, all related to the cubic perovskite structure by the freezing-in of a 
polar mode at $\Gamma$.
The main features of the phonon dispersion of 
BaTiO$_3$ have been previously discussed in Ref. \cite{Ghosez98a} and are very 
similar to those of KNbO$_3$~\cite{Yu95b}.
The most unstable mode is at $\Gamma$, and this mode, dominated 
by the Ti displacement against the oxygens (Table~\ref{Table.w}), is 
the one that freezes in to give the ferroelectric phases. However, 
the instability is not restricted to the $\Gamma$ point. Branches 
of Ti-dominated unstable modes extend over much of the  
Brillouin zone. The flat dispersions of the unstable transverse optic
mode towards X and M, combined with its rapid stiffening towards R, 
confine the instability to three quasi-two-dimensional ``slabs" of reciprocal 
space intersecting at $\Gamma$.
This is the fingerprint of a ``chain-like'' unstable localized distortion for 
the Ti displacements in real space~\cite{Yu95b,Ghosez98a}. Except for these modes,
all the other phonons are stable in BaTiO$_3$, which makes the behavior of the 
unstable branches relatively easy to understand.

Lead titanate (PbTiO$_3$) 
has a single transition to a low-temperature ferroelectric tetragonal structure, 
related to the cubic perovskite structure by the freezing-in of a polar mode at $\Gamma$.
The phonon dispersion of PbTiO$_3$ shows similar features to that of BaTiO$_3$, 
with some important differences.
As in BaTiO$_3$, the most unstable mode is at $\Gamma$, consistent with the  
observed ground state structure. However, the 
eigenvector is no longer strongly dominated by the displacement of  the
Ti against the oxygen along the Ti--O chains, but contains a 
significant component of the Pb moving against the O atoms in the Pb--O 
planes (see Table~\ref{Table.w}). Unstable 
Ti-dominated modes, similar to those in BaTiO$_3$, 
can be identified in the vicinity of the M--X line 
(M$_{3'}$, X$_5$ modes). However, Pb now plays an active 
role in the character of the majority of the unstable branches, notably those terminating at 
M$_{5'}$ and X$_{5'}$. Also, the Pb-dominated branch emanating from the ferroelectric 
$\Gamma$ mode towards R has a much weaker 
dispersion than the corresponding, Ti-dominated, branch in BaTiO$_3$. 
In consequence, the unstable localized ferroelectric distortion in real 
space is nearly isotropic, in contrast to the pronounced anisotropy in
BaTiO$_3$. 
Finally, there is an antiferrodistortive instability at the 
R-point (R$_{25}$ mode). As similarly observed in 
SrTiO$_3$~\cite{LaSota97}, 
this instability is confined to
quasi-one-dimensional ``tubes" of reciprocal space running along the edges
of the simple cubic Brillouin zone (R$_{25}$ and M$_3$ modes and the branch
connecting them).  The branches emanating from this region stabilize rapidly 
away from the Brillouin zone edge towards, in particular, $\Gamma_{25}$ and X$_3$.
In real space, this instability appears as a cooperative rotation of oxygen 
octahedra, with strong correlations in the plane perperpendicular to the axis of
rotation, and little correlation between rotations in different planes. 
The lack of interplane correlation, arising from the flatness of the R$_{25}$--M$_3$ 
branch, suggests the absence of coupling between the oxygen motion in different planes. 
This will be discussed further in the next section.

The ground state of PbZrO$_3$ is an antiferroelectric with 8 formula units per unit cell, 
obtained by freezing in a set of coupled modes, most importantly modes at R 
and $\Sigma(\frac{1}{4}\frac{1}{4}0)$\cite{Fujishita84}.
The phonon dispersion correspondingly shows even more pronounced and complex 
instabilities than for PbTiO$_3$. 
Overall, the unstable branches are dominated by Pb and O displacements, with  
no significant Zr character. There is still a polar instability at the 
$\Gamma$ point but the eigenvector (see  Table~\ref{Table.w}) 
is clearly dominated by the  displacement of lead against the oxygens while 
the Zr atom now moves {\em with} these oxygens. In fact, the modes 
where the Zr is displaced against the oxygens ($\Gamma_{\rm LO}$ at 160 cm$^{-1}$, 
M$_{3'}$, 
X$_5$ modes) are now all stable. The octahedral rotation branch is again 
remarkably flat and is significantly more unstable at R$_{25}$ and M$_3$ than in 
PbTiO$_3$. The antiferrodistortive instability retains some 
one-dimensional character but spreads into a larger region of reciprocal space~: 
the $\Gamma_{25}$ and X$_3$ transverse oxygen motions, related to the R$_{25}$ mode, 
are still stable but with a relatively low frequency. 
We note finally that the stiffest longitudinal and tranverse oxygen branches
have been shifted to higher energy relative to the titanates.

\section{Interatomic force constants}

In the previous section, comparisons between the three compounds were made by  
analyzing phonon dispersion relations along high-symmetry lines in reciprocal  
space.
A complementary, highly instructive picture of the quadratic-order structural  
energetics of the system is provided by 
direct examination of the real-space interatomic force constants (IFC).

The interatomic force constants (IFC) are  
generated in the construction of the phonon dispersion relations; their computation 
has been described in Section II. 
Our convention 
is that the IFC matrix $C_{\alpha,\beta}(l\kappa,l'\kappa')$ which relates
the force $F_{\alpha}(l\kappa)$ on atom $\kappa$ in cell $l$ and 
the displacement $\Delta \tau_{\beta}(l'\kappa')$ of atom $\kappa'$ in cell  
$l'$ is defined through the following expression: 
$F_{\alpha}(l\kappa) = - C_{\alpha,\beta}(l\kappa,l'\kappa') . \Delta  
\tau_{\beta}(l'\kappa')$.
Moreover, the total IFC can be decomposed into a dipole-dipole part (DD) 
and a short-range part (SR) , following Refs. ~\cite{Gonze94,Ghosez97}. 
Such a decomposition is somewhat arbitrary but is 
useful for understanding the microscopic origin of the trends among
different compounds. For convenience, the atoms are labeled according 
to Table~\ref{Table.pos}, as illustrated in Fig.~\ref{Fig.AT}. 
The interatomic force
constants are reported either in cartesian
coordinates or in terms of their longitudinal
($\parallel$) and transverse ($\perp$)
contributions along the line connecting the two atoms. The results for BaTiO$_3$, 
PbTiO$_3$ and PbZrO$_3$ 
are presented in Tables \ref{Table.IFC-SF}, \ref{Table.IFC-A} 
and \ref{Table.IFC-B}.

First, we examine the ``self-force constant,'' which specifies the force on a single 
isolated atom at a unit displacement from its crystalline position, 
all the other atoms remaining fixed. The values are given in Table  
\ref{Table.IFC-SF}. The self-force constants are positive for all atoms in the three  
compounds, so that all three are stable against isolated atomic displacements. 
Therefore, it is only the cooperative motion of different atoms that can  
decrease 
the energy of the crystal and generate an instability, such as is observed in  
the phonon dispersion relations presented in the previous Section.  The  
analysis of the IFCs will
help us to identify the energetically favorable coupling in the displacements  
and elucidate the origin of the unstable phonon branches.

Next, we discuss the ferroelectric instability at $\Gamma$, and the phonon branches 
which emanate from it. 
In barium titanate, it was found that the unstable eigenvector is dominated by Ti 
displacement along the Ti--O--Ti chain.
If we consider the simple case where only Ti atoms are allowed to displace, we find 
that the destabilizing contribution from the Ti$_0$--Ti$_1$ $\parallel$ interaction 
itself is nearly enough to compensate the Ti self-force constant 
(Table \ref{Table.IFC-A}). 
In addition, the fact that the Ti$_0$--Ti$_1$ $\perp$ interaction is comparatively small 
can account directly for the characteristic flat dispersion 
along $\Gamma$-X and $\Gamma$-M 
and the strong stiffening along $\Gamma$-R, associated with the chain-like nature of 
the instability.
For the true eigenvector, another important, though relatively small, destabilizing 
contribution comes from the cooperative displacement 
of the O$_1$ atoms against the titaniums along the Ti--O chains. This, together with 
the total contribution of the rest of the IFCs, is responsible for the actual instability 
of the ferroelectric Ti-dominated branches in BaTiO$_3$. 

For lead titanate, the energetics of the Ti-only displacements, dominated by the Ti 
self-force constant and the Ti$_0$--Ti$_1$ $\parallel$ and $\perp$ interactions, are 
remarkably similar to those in BaTiO$_3$ (Table \ref{Table.IFC-A}). 
However, in PbTiO$_3$ there is also an important destabilization associated with pure 
Pb displacements~\cite{Rem-Pb}. This can be fully attributed to the large difference 
in the Ba and Pb self-force constants, 
while the  A$_0$--A$_1$ $\parallel$ and $\perp$ interactions are very similar 
in the two compounds. 
Also, the  A$_0$--B$_0$ $\parallel$ and $\perp$ cation interactions are of the same 
order of magnitude as in BaTiO$_3$ and combine to give a surprisingly small
$xx$ coupling.
At $\Gamma$, symmetry considerations permit the mixing of Ti--O and Pb--O displacements 
and in the phonon branches which emanate from it, thus accounting for the nature 
of the ferroelectric eigenvector. 
However, at X, M and R symmetry labels distinguish the Ti-dominated (X$_5$, M$_{3'}$ and 
R$_{25'}$)  and Pb-dominated (X$_{5'}$, M$_{2'}$ and R$_{15}$) modes, which can be readily 
identified in the calculated phonon dispersion.
Also, the Pb$_0$--Pb$_1$ 
coupling is much smaller in magnitude than the Ti$_0$--Ti$_1$ coupling, which accounts for 
the relatively weak dispersion of the Pb-dominated branch from $\Gamma$ to R.
In the true eigenvectors, these instabilities are further reinforced
by displacements of the oxygens. While the longitudinal IFC between Ba$_0$
and O$_1$ was very small in BaTiO$_3$,  there is a significant destabilizing
interaction between Pb$_0$ and O$_1$  in PbTiO$_3$, which further promotes
the involvement of Pb in the unstable phonon branches.
We note that the Ti$_0$--O$_1$ longitudinal interaction is repulsive in PbTiO$_3
$,
but it is even smaller in amplitude than in BaTiO$_3$ and its stabilizing effect
 is
compensated by the transverse coupling between Pb and O$_1$.

In lead zirconate, the unstable eigenvector at $\Gamma$ is strongly dominated by 
Pb--O motion, with little involvement of Zr. 
This can be understood by comparing, in Table \ref{Table.IFC-A}, 
the energetics of Zr-only displacements with those of Ti-only displacements 
in PbTiO$_3$ and BaTiO$_3$: the Zr self-force constant is significantly 
larger and the Zr$_0$--Zr$_1$ $\parallel$ and $\perp$ interactions are smaller, so that 
Zr cannot move as easily as Ti.
Also,  the Zr$_0$--O$_ 1$ $\parallel$ interaction is now significantly repulsive, 
explaining why the Zr atom does not move against the oxygens, but with them. 
As for the titanates, we note 
finally that the Zr atoms are mainly coupled along the B--O chains, so that the 
characteristic dispersion of the B-atom modes is preserved, only at higher frequencies.
On the other hand, the Pb self-force constant is much smaller, the Pb$_0$--Pb$_1$ 
$\parallel$ 
and $\perp$ interactions are only slightly smaller, and the destabilizing coupling 
between lead and oxygen  
is similar to that in PbTiO$_3$, accounting for the involvement of Pb in the 
instability.

Finally, we discuss the antiferrodistortive instability 
identified with the R$_{25}$ and M$_3$ modes and the branch along R--M connecting them.
There is a marked variation in the frequency of the R$_{25}$ mode in the three compounds, 
ranging from the lack of any instability
in BaTiO$_3$, to PbTiO$_3$ with an unstable R$_{25}$ mode that nonetheless 
does not contribute to the ground state, and finally to PbZrO$_3$ in which the
R$_{25}$ mode is even more unstable and contributes significantly to
the observed ground state \cite{Fujishita84}.
The eigenvector of this mode is completely determined by 
symmetry and corresponds to a coupled rotation 
of the corner-connected oxygen octahedra. Its  
frequency depends only on the oxygen IFCs, predominantly the self-force constant and the 
off-diagonal coupling
between nearest neighbor oxygen atoms. 
In fact, the latter (for example, O$_{1y}$--O$_{2z}$ in Table \ref{Table.IFC-B}) 
is remarkably 
similar in all three compounds. The trend is therefore associated with the rapid 
decrease in the transverse O self-force constant from BaTiO$_3$ to PbTiO$_3$ to PbZrO$_3$ 
and the resulting compensation of the contribution from the self-force constant by the 
destabilizing contribution from the off-diagonal coupling.

The self-force constant can be written as a sum over interatomic force constants, 
according to the requirement of translational 
invariance: $C_{\alpha,\beta}(l\kappa,l\kappa) =  
-\sum_{l'\kappa'}'
C_{\alpha,\beta}(l\kappa,l'\kappa')$. 
It is therefore of interest to identify which interatomic force constants are 
responsible for the trend in the transverse oxygen self-force constant. 
The suggestion that the trend is due to covalency-induced changes in the 
Pb--O interactions can be directly investigated through a ``computer experiment.''
Everything else being equal, we artificially replace the IFC between A$_0$ 
and O$_1$ atoms in BaTiO$_3$ by 
its value in PbTiO$_3$, consequently modifying the self-force constant 
on A and O atoms. For this hypothetical material, the A-atom dominated modes are  
shifted to lower frequencies while the frequency of the R$_{25}$ mode 
is lowered to 40$i$ cm$^{-1}$. If we introduce the stronger A$_0$--O$_1$  
interaction of PbZrO$_3$, we obtain an even larger R$_{25}$ instability 
of 103$i$ cm$^{-1}$. 

The previous simulation demonstrates the crucial role played by the 
lead-oxygen interaction in generating the AFD instability. However, 
this change alone is not sufficient to reproduce
the flatness of the R$_{25}$--M$_3$ branch, as the corresponding frequencies 
of the M$_3$ mode in the two hypothetical cases above are 92 cm$^{-1}$ 
and 25$i$ cm$^{-1}$, respectively. 
Naively, the absence of dispersion 
of the antiferrodistortive mode along that line would be interpreted
as the absence of coupling between the oxygens in the different 
planes. 
However, as can be seen in Table \ref{Table.IFC-B}, 
the $yy$ transverse coupling between O$_1$ and O$_3$ 
is far from negligible, and acts to amplify the AFD instability at R with respect to M.
In the lead compounds, however, this is compensated  by another $yz$ coupling, between 
O$_1$ and O$_5$.  
The latter is significantly smaller in BaTiO$_3$ (by 35 \%). If we consider a third 
hypothetical compound in which this coupling in BaTiO$_3$ is additionally 
changed to its value in 
PbTiO$_3$, we recover a flat behavior along the R--M line. 
In the lead perovskites, the flatness of this band appears therefore as a consequence 
of a compensation between different interplane interactions, 
and cannot be attributed to complete independence
of oxygen motions in the different planes.

\section{Discussion}

In Section IV, we observed marked differences between the phonon dispersion 
relations and eigenvectors in the three related compounds.
Through the real-space analysis in the previous section, we have seen that 
these differences arise from changes in a few key interatomic force constants.

First, we remark that B--O interactions depend strongly on the B atom, being 
similar in PbTiO$_3$ and BaTiO$_3$, and quite different in PbZrO$_3$.
In fact, the SR force contribution to the Zr$_0$--O$_1$ 
interaction and Ti$_0$--O$_1$ are very similar, so that the difference arises 
from the dipolar contribution.
In PbZrO$_3$, this contribution is reduced in consequence of the lower values 
of the Born effective charges (see Table \ref{Table.IFC-A}).
This trend provides another example of the 
{\it very} delicate nature of the compensation between SR and DD forces, 
previously pointed out for BaTiO$_3$~\cite{Ghosez96,Ghosez97}.

Next, we remark that A--O interactions depend strongly on the A atom, being 
similar in PbTiO$_3$ and PbZrO$_3$, and quite different in BaTiO$_3$.
This change originates in 
the covalent character of the bonding between Pb and O, which results both 
in smaller A--O SR coupling and a larger Born effective charge for Pb.
Even though the impact of the latter on destabilizing the DD interaction 
is partly compensated by the increased $\epsilon_{\infty}$, the net effect 
is to promote the Pb--O instability.

As discussed above, the self-force constant can be written as a sum over 
interatomic force constants.
It can be easily verified that the trends in the self-force constants observed 
in Table III are primarily associated with the trends in A--O and B--O interactions.

The rest of the IFCs given in Table VI are actually remarkably similar.
For example, A--B interactions are apparently insensitive to the identity of A (Ba, Pb) 
or B (Ti, Zr). 
This is true also for A--A, B--B and most O--O interactions. 
The small differences observed can at least in part be attributed to differences in 
the lattice constants and in $\epsilon_{\infty}$ for the three compounds.

The similarities in IFC's among compounds with related compositions offer an 
intriguing opportunity for the modelling of the lattice dynamics of solid solutions. 
In the simplest case, the lattice dynamics of ordered supercells of  compounds such as 
PZT could be obtained by using the appropriate A--O and B--O couplings from the pure 
compounds and averaged values for the A--B, A--A, B--B and O--O interactions.
In a more sophisticated treatment, the dipolar contributions could be separately handled 
within an approach correctly treating the local fields and Born effective charges.
Implementation of these ideas is in progress.

\section{Conclusions}

In this paper, we have described in detail the first-principles phonon dispersion 
relations and real-space interatomic force constants  
of cubic PbTiO$_3$ and PbZrO$_3$ and compared them with results previously obtained 
for BaTiO$_3$. 
The modifications induced by the substitution
of Ba by Pb and of Ti by Zr are seen to be most easily understood by considering the 
real-space IFCs.
The replacement of Ba by Pb strongly strengthens the A--O coupling, which is directly 
responsible for both the involvement of Pb in the ferroelectric eigenvector and the 
appearance of the antiferrodistortive instability.
The two-dimensional real-space character of the latter results from 
an additional slight modification of the O--O coupling. 
The replacement of Ti in PbTiO$_3$
by Zr strongly 
modifies the B--O interaction, suppressing the involvement of Zr in the 
unstable modes of PbZrO$_3$.
The decrease of the Born effective charges along the B--O bonds
is a crucial factor in modifying this interaction.
In addition, the substitution of Ti by Zr slightly strengthens the Pb--O  
coupling. 
Apart from these modifications,  the other IFCs are remarkably similar in the three 
compounds studied.
The consequent prospects for transferability to solid solutions were discussed.

\acknowledgments
The authors thank Ch. LaSota and H. Krakauer for communicating unpublished   
results concerning SrTiO$_3$ as well as X. Gonze for the availability of 
the {\sc abinit} package, helpful for the interpolation of the phonon 
dispersion curves and the analysis of the interatomic force constants. 
PhG, UVW and KMR acknowledge useful discussions with many participants in 
the 1998 summer workshop ``The Physics of Insulators'' at the Aspen Center 
for Physics, where part of this work was performed. This work was supported 
by ONR Grant N00014-97-0047. EC received partial funding from an NRC 
Postdoctoral Fellowship and from Sandia 
National Laboratories.  Sandia is a multi-program national laboratory
operated by the Sandia Corporation, a Lockheed Martin Company,
for the United States Department of Energy under contract
DE-AL04-94AL8500.

\end{multicols}


\begin{figure}
\caption{Calculated phonon dispersion relations of BaTiO$_3$, PbTiO$_3$ and  
PbZrO$_3$ along various high-symmetry lines in the simple cubic Brillouin zone. 
A color has been assigned to each point based on the contribution of each 
kind of atom to the associated dynamical matrix eigenvector
(red for the A atom, green for the B atom, and blue for the
oxygens) \protect\cite{Note-color}.
Symmetry labels follow the convention of Ref. \protect\onlinecite{Cowley}, 
with the A atom at the origin.}
\label{Fig.BS}
\end{figure}

\begin{figure}
\caption{Schematic three-dimensional view of the atoms labeled
in Table \protect \ref{Table.pos}.}
\label{Fig.AT}
\end{figure}


\begin{table}
\caption {Lattice parameter (\AA) at which DFPT calculations were performed, 
Born effective charges ($|e|$) and 
optical dielectric 
constant in the cubic perovskite structure for the three ABO$_3$ compounds.}
\begin {tabular}{lccc}
 $~$   &BaTiO$_3$    &PbTiO$_3$   &PbZrO$_3$ \\
\hline
a$_{cell}$                    &$4.00$      &$3.97$       &$4.12$    \\
$Z^{*}_{A}$                   &$+2.74$     &$+3.87$      &$+3.93$   \\
$Z^{*}_{B} $                  &$+7.32$     &$+7.04$      &$+5.89$   \\
$Z^{*}_{O_{\perp}}$           &$-2.14$     &$-2.57$      &$-2.50$   \\
$Z^{*}_{O_{\parallel}}$       &$-5.78$     &$-5.76$      &$-4.82$   \\
$\epsilon_{\infty}$           &$ 6.75$     &$ 8.24$      &$ 6.97$   \\
\end{tabular}
\label{Table.Z}
\end{table}

\begin{table}
\caption {Normalized dynamical matrix eigenvector for the unstable 
ferroelectric mode at $\Gamma$ (z-polarization). The corresponding
eigendisplacement in real space can be obtained by dividing each 
value by the appropriate mass factor $\protect\sqrt{M_{ion}}$ .
}
\begin {tabular}{lccccc}
ABO$_3$   &A  &B   &O$_x$  &O$_y$    &O$_z$ \\
\hline
BaTiO$_3$  &$+0.0178$  &$+0.6631$  &$-0.2842$  &$-0.2842$  &$-0.6311$ \\  
PbTiO$_3$  &$+0.2314$  &$+0.4024$  &$-0.4792$  &$-0.4792$  &$-0.5704$ \\
PbZrO$_3$  &$+0.5033$  &$-0.1786$  &$-0.5738$  &$-0.5738$  &$-0.2374$ \\
\end{tabular}
\label{Table.w}
\end{table}

\begin{table}
\caption {Label assigned to various atoms in terms of their position in
reduced coordinates.}
\begin {tabular}{cccccc}
A$_0$ & ( 0.0, 0.0, 0.0) & B$_0$ & ( 0.5, 0.5, 0.5) &O$_1$ & ( 0.5, 0.5, 0.0) \\
A$_1$ & ( 0.0, 0.0, 1.0) &B$_1$ & ( 1.5, 0.5, 0.5) & O$_2$ & ( 0.5, 0.0, 0.5)   \\
& & & &O$_3$  & (-0.5, 0.5, 0.0) \\
& & & &O$_4$ & ( 0.5, 0.5,-1.0) \\
& & & &O$_5$ & (-0.5, 0.0, 0.5)    \\
\end{tabular}
\label{Table.pos}
\end{table}

\begin{table}
\caption {Self-force constant (Ha/Bohr$^2$) on the different atoms in the 
unit cell. }
\begin {tabular}{lcccc}
Atom   &Direction  &{BaTiO$_3$}  &{PbTiO$_3$}  &{PbZrO$_3$} \\
\hline
A$_0$  & x=y=z    &$+0.0806$   &$+0.0247$  &$+0.0129$  \\
B$_0$  & x=y=z    &$+0.1522$   &$+0.1393$  &$+0.2302$  \\
O$_1$  & x=y      &$+0.0681$   &$+0.0451$  &$+0.0166$  \\ 
       & z        &$+0.1274$   &$+0.1518$  &$+0.2758$  \\  
\hline
\end{tabular}
\label{Table.IFC-SF}
\end{table}

\begin{table}
\caption {Selected longitudinal ($\parallel$), transverse ($\perp$) 
and cartesian (${\alpha \beta}$) interatomic force constants
(Ha/Bohr$^2$) between different pairs of atoms. The dipole-dipole (DD) and  
remaining 
short-range (SR) contribution, have been separated following the scheme  
described in
Ref.~\protect\cite{Ghosez98a}.}
\begin {tabular}{lcccccccccc}
& &\multicolumn{3}{c}{BaTiO$_3$}  &\multicolumn{3}{c}{PbTiO$_3$}  
&\multicolumn{3}{c}{PbZrO$_3$} \\
Atom  & &Total     &DD    &SR  &Total     &DD    &SR  &Total     &DD    &SR \\
\hline
B$_0$-O$_1$ &(${\parallel}$)    &$+0.0094$     &$+0.2325$      &$-0.2231$        
                                &$-0.0012$     &$+0.1865$      &$-0.1877$        
                                &$-0.0687$     &$+0.1380$      &$-0.2067$         
\\
            &(${\perp}$)        &$-0.0211$     &$-0.0430$      &$+0.0218$        
                                &$-0.0178$     &$-0.0417$      &$+0.0239$        
                                &$-0.0100$     &$-0.0358$      &$+0.0258$         
\\
B$_0$-B$_1$ &(${\parallel}$)    &$-0.0672$     &$-0.0368$      &$-0.0304$     
                                &$-0.0615$     &$-0.0285$      &$-0.0330$        
                                &$-0.0499$     &$-0.0211$      &$-0.0288$         
\\
            &(${\perp}$)        &$+0.0075$     &$+0.0184$      &$-0.0109$    
                                &$+0.0065$     &$+0.0142$      &$-0.0077$        
                                &$+0.0054$     &$+0.0105$      &$-0.0052$         
\\
B$_0$-O$_4$ &(${\parallel}$)    &$+0.0156$     &$+0.0086$      &$+0.0070$        
                                &$+0.0135$     &$+0.0069$      &$+0.0066$                                       &$+0.0106$     &$+0.0051$      &$+0.0055$         
\\
            &(${\perp}$)        &$+0.0009$     &$-0.0016$      &$+0.0007$    
                                &$+0.0015$     &$-0.0015$      &$+0.0006$        
                                &$+0.0012$     &$-0.0013$      &$+0.0002$         
\\                                

B$_0$-A$_0$ &({$\parallel$})    &$-0.0286$     &$-0.0212$      &$-0.0074$    
                                &$-0.0277$     &$-0.0241$      &$-0.0036$        
                                &$-0.0271$     &$-0.0216$      &$-0.0054$         
\\ 
            &({$\perp$})        &$+0.0134$     &$+0.0106$      &$+0.0028$    
                                &$+0.0157$     &$+0.0121$      &$+0.0036$        
                                &$+0.0145$     &$+0.0108$      &$+0.0037$         
\\ 
            &(${xx}$)           &$-0.0006$     &$+0.0000$      &$-0.0006$    
                                &$+0.0012$     &$+0.0000$      &$+0.0012$        
                                &$+0.0007$     &$+0.0000$      &$+0.0007$         
\\                                
A$_0$-O$_1$ &(${\parallel}$)    &$-0.0004$     &$+0.0114$      &$-0.0118$        
                                &$+0.0108$     &$+0.0162$      &$-0.0054$        
                                &$+0.0139$     &$+0.0169$      &$-0.0030$         
\\
              &(${zz}$)         &$-0.0108$     &$-0.0154$      &$+0.0045$
                                &$-0.0110$     &$-0.0181$      &$+0.0071$
                                &$-0.0103$     &$-0.0163$      &$+0.0060$
\\
A$_0$-A$_1$ &(${\parallel}$)    &$-0.0112$     &$-0.0052$      &$-0.0060$     
                                &$-0.0108$     &$-0.0086$      &$-0.0022$        
                                &$-0.0094$     &$-0.0093$      &$-0.0001$         
\\
                &(${\perp}$)    &$+0.0038$     &$+0.0025$      &$+0.0012$     
                                &$+0.0054$     &$+0.0043$      &$+0.0011$        
                                &$+0.0056$     &$+0.0047$      &$+0.0009$         
\\
\end{tabular}
\label{Table.IFC-A}
\end{table}

\begin{table}
\caption {Interatomic force constant matrix in cartesian coordinates
(Ha/Bohr$^2$) between various pairs of oxygen atoms. Lines and columns of
the matrix correspond respectively to x, y and z displacements for the 
first and second atom mentioned in the first column of the Table.}
\begin {tabular}{lccc}
Atoms &{BaTiO$_3$}  &{PbTiO$_3$}  &{PbZrO$_3$} \\
\hline
\\
O$_1$-O$_2$ &$\left( \begin{array}{rrr} +0.0037 &0.0000 &0.0000 \\ 
                                        0.0000 &-0.0087 &+0.0119 \\ 
                                        0.0000 &+0.0274 &-0.0087 \\ \end{array}  \right) $ 
            &$\left( \begin{array}{rrr} +0.0035 &0.0000 &0.0000 \\ 
                                        0.0000 &-0.0091 &+0.0123 \\ 
                                        0.0000 &+0.0271 &-0.0091 \\ \end{array}  \right) $ 
            &$\left( \begin{array}{rrr} +0.0038 &0.0000 &0.0000 \\ 
                                        0.0000 &-0.0065 &+0.0110 \\ 
                                        0.0000 &+0.0229 &-0.0065 \\ \end{array}  \right) $ \\
\\
O$_1$-O$_3$ &$\left( \begin{array}{rrr} -0.0019 &0.0000 &0.0000 \\ 
                                         0.0000 &+0.0017 &0.0000 \\ 
                                         0.0000 &0.0000 &+0.0091 \\ \end{array}  \right) $ 
            &$\left( \begin{array}{rrr} -0.0012 &0.0000 &0.0000 \\ 
                                         0.0000 &+0.0022 &0.0000 \\ 
                                         0.0000 &0.0000 &+0.0079 \\ \end{array}  \right) $ 
            &$\left( \begin{array}{rrr} -0.0012 &0.0000 &0.0000 \\ 
                                         0.0000 &+0.0021 &0.0000 \\ 
                                         0.0000 &0.0000 &+0.0055 \\ \end{array}  \right) $ \\
\\
O$_1$-O$_4$ &$\left( \begin{array}{rrr} -0.0003 &0.0000 &0.0000 \\ 
                                         0.0000 &-0.0003 &0.0000 \\ 
                                         0.0000 &0.0000 &-0.0321 \\ \end{array}  
\right) $ 
            &$\left( \begin{array}{rrr}  +0.0003 & 0.0000 &0.0000 \\ 
                                         0.0000 & +0.0003 &0.0000 \\ 
                                         0.0000 & 0.0000 &-0.0326 \\  
\end{array} \right) $ 
            &$\left( \begin{array}{rrr} -0.0010 & 0.0000 &0.0000 \\ 
                                         0.0000 &-0.0010 &0.0000 \\ 
                                         0.0000 & 0.0000 &-0.0362 \\  
\end{array} \right) $ \\
\\
O$_1$-O$_5$ &$\left( \begin{array}{rrr}  -0.0006 &-0.0013 &+0.0007 \\ 
                                         -0.0007 &+0.0013 &+0.0007 \\ 
                                         +0.0013 &+0.0025 &+0.0013 \\ \end{array}  
\right) $ 
            &$\left( \begin{array}{rrr}  -0.0010 &-0.0013 &+0.0010 \\ 
                                         -0.0010 &+0.0012 &+0.0011 \\ 
                                         +0.0013 &+0.0022 &+0.0012 \\ \end{array}  
\right) $ 
            &$\left( \begin{array}{rrr}  -0.0010 &-0.0013 &+0.0010 \\ 
                                         -0.0010 &+0.0011 &+0.0010 \\ 
                                         +0.0013 &+0.0018 &+0.0011 \\ \end{array}  
\right) $ \\

\end{tabular}
\label{Table.IFC-B}
\end{table}


\begin{thebibliography}{10}
\bibitem{Rem-XC}
In our version of the {\sc castep} code, the XC energy density ($\epsilon_{xc}$) and 
XC potential ($v_{xc}$) are first evaluated on a grid from the Perdew-Zunger 
analytical expression. During the computation, the values of $\epsilon_{xc}$ 
and $v_{xc}$ at a given $r_s$ are then interpolated using 
to a third-order polynomial fit of the grid values in the region of interest
(i.e. using the 4 grid points closest to the given $r_s$). 
In our DFPT calculation, the exchange part of the XC kernel 
($K_{xc}$) has been treated exactly while the correlation contribution was 
evaluated from the first derivative of a third-order polynomial fit of the grid values 
of the XC potential. This technique, by smoothing out the $v_{xc}$ curve, 
eliminates the problem associated with the discontinuity at $r_s=1$ of the second 
derivative of Perdew-Zunger exchange-correlation energy analytical expression.
The resulting difference
between force constants determined from frozen phonon
and DFPT calculations is negligible.
For example, using this technique in PbTiO$_3$ gives the diagonal matrix element 
at $\Gamma$
associated with the displacement of oxygen in the Pb--O plane as 3.3337 eV/\AA$^2$, 
while the frozen phonon value is equal to 3.3328 eV/\AA$^2$. 
Further details will be described in a separate publication.
\bibitem{Note-mesh}
Discrete Fourier transform from a $2 \times 2 \times 2$ simple cubic (SC) mesh 
of $q$-points would limit the range of the SR part of the IFC to 
a $2 \times 2 \times 2$ supercell in real space. Our BCC mesh summarizes the
information of two SC meshes shifted with respect to each other by a vector
$q=$(0.25, 0.25, 0.25).
\bibitem{Note-color}
For example, a normalized mode with A displacement 0.7, 
B displacement 0.5, and O displacements 0.5, 0.1, and 
0.0 (in generalized coordinates) would be colored via 
the command ``0.49 0.25 0.26 setrgbcolor" in the Adobe 
Systems Inc. PostScript$^{\rm TM}$ language.
\bibitem{Rem-Pb}
Specifically, the Pb diagonal element of the full force constant matrix 
at $\Gamma$ ($-0.0018$ Ha/Bohr$^2$) shows that uniform Pb displacement is 
unstable. This is not the case for Ba in BaTiO$_3$.
\end{thebibliography}
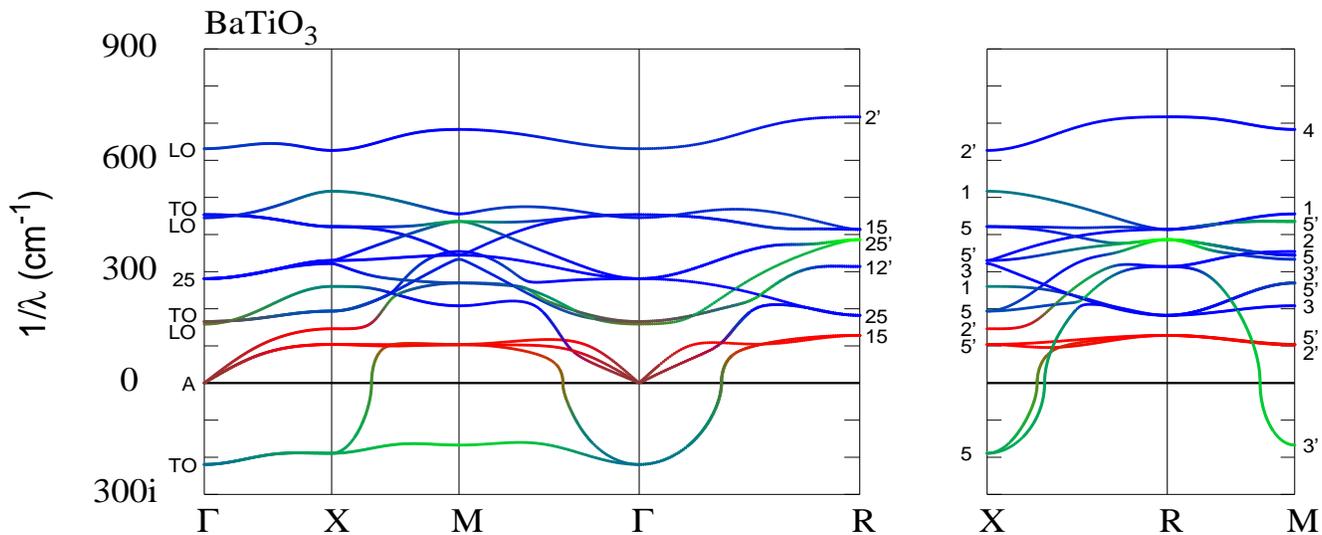
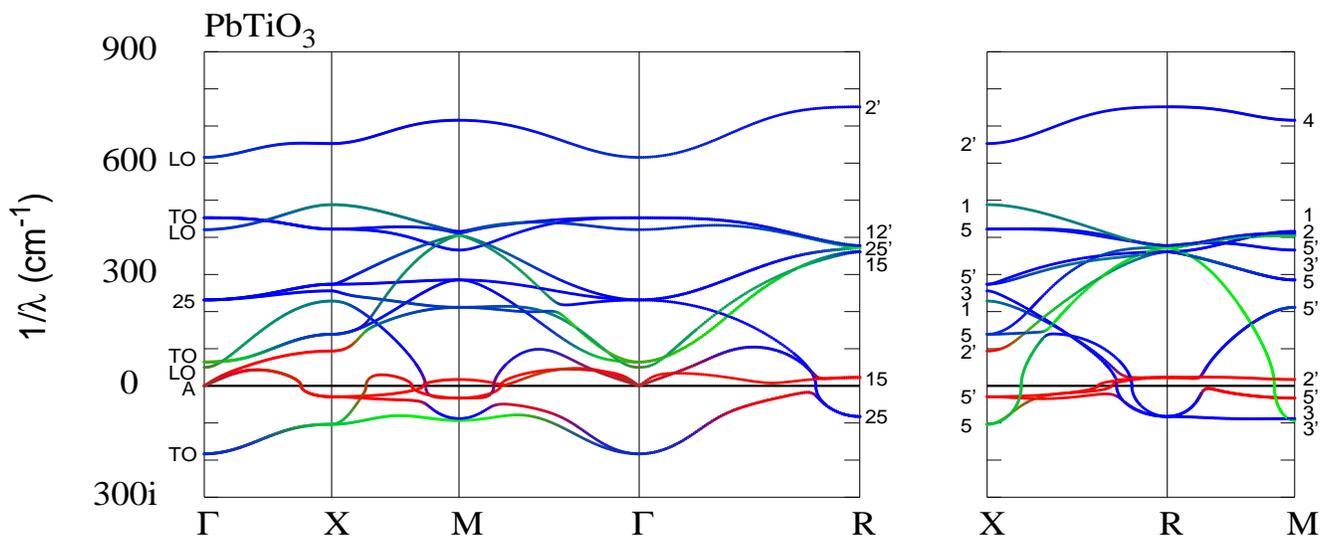
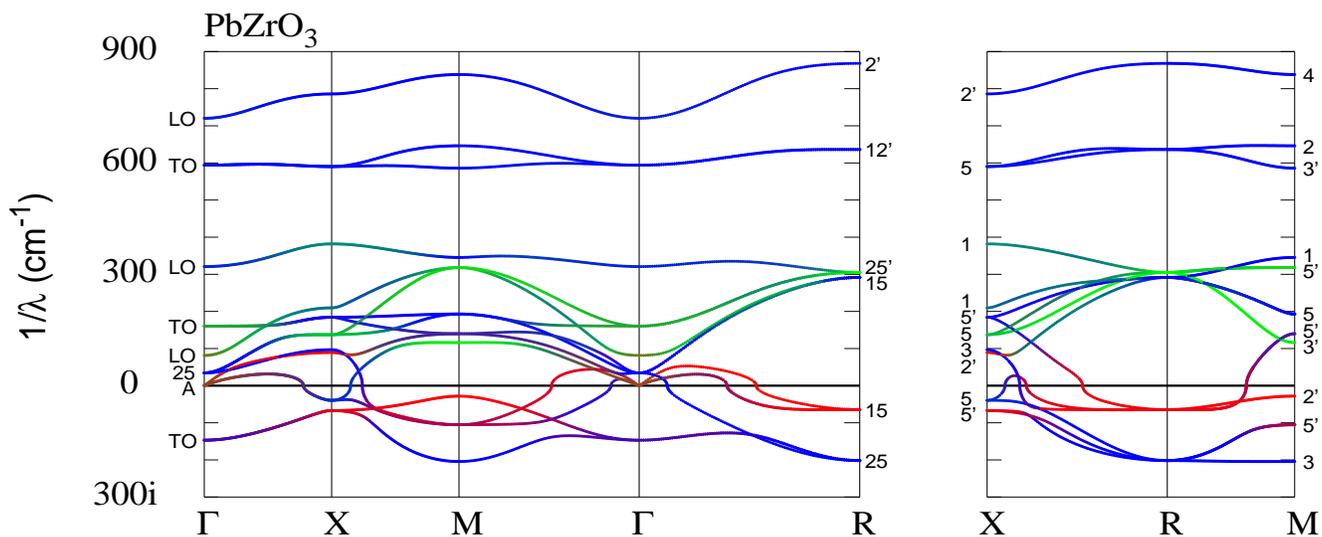
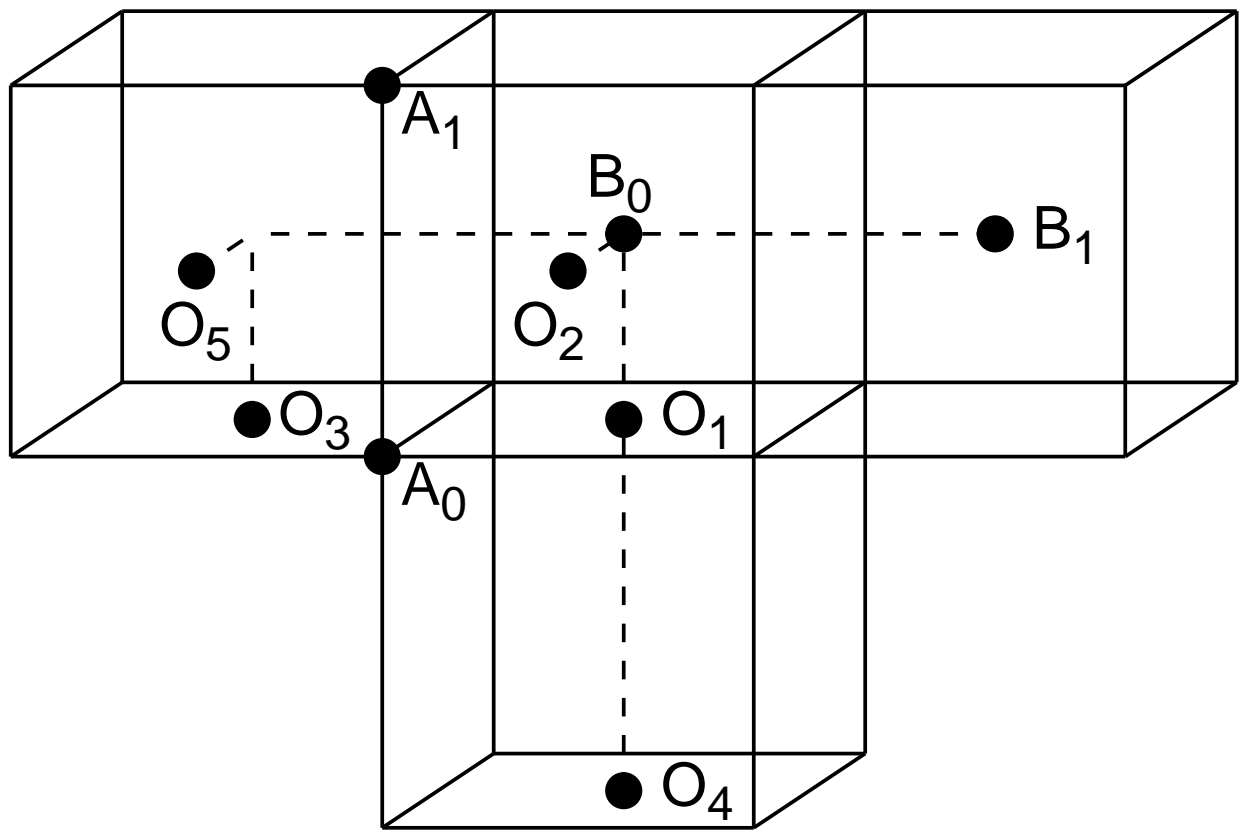
\end{document}